# Telepathology in Hematopathology Diagnostics: A Collaboration Between Ho Chi Minh City Oncology Hospital and University of Texas Health-McGovern Medical School

Uyen Ly[1], MD; Quang Nguyen[1], MD; Dang Nguyen[1], MD; Tu Thai[1], MD; Binh Le[1], MD; Duong Gion[1], MD; Alexander Banerjee[2], DO; Brenda Mai[2], MD; Amer Wahed[2], MD; Andy Nguyen[2]*, MD
 [1] Pathology Department, Ho Chi Minh City Oncology Hospital, Vietnam
 [2] Pathology Department, University of Texas Health-McGovern Medical School, Houston, USA

---

**ABSTRACT**
Digital pathology in the form of whole-slide-imaging has been used to support diagnostic consultation through telepathology. Previous studies have mostly addressed the technical aspects of telepathology and general pathology consultation. In this study, we focus on our experience at University of Texas Health-McGovern Medical School in Houston, Texas in providing hematopathology consultation to the Pathology Department at Ho Chi Minh City Oncology Hospital in Vietnam. Over a 32-month period, 71 hematopathology cases were submitted for telepathology. Diagnostic efficiency significantly improved with average turnaround times reduced by 30% compared to traditional on-site consultations with local pathologists using glass slides. A web site has been established in this telepathology project to retain information of the most recently discussed cases for further review after the teleconference. Telepathology provides an effective platform for real-time consultations, allowing remote subspecialty experts to interact with local pathologists for comprehensive case reviews. This process also fosters ongoing education, facilitating knowledge transfer in regions where specialized hematopathology expertise is limited.

---

*Corresponding author
Andy Nguyen, MD
Department of Pathology and Laboratory Medicine
University of Texas Health Science Center at Houston
McGovern Medical School
6431 Fannin Street, MSB 2.292, Houston, TX, 77030
(713) 500-5337; fax (713) 500-0712
Nghia.D.Nguyen@uth.tmc.edu

**INTRODUCTION**
Telepathology, a transformative extension of traditional pathology, enables remote diagnostic evaluation by transmitting microscopic images. This technology facilitates critical applications in diagnostics, second opinions, and various educational and research initiatives [1]. Digital slides, rather than traditional glass slides, are transmitted for review, enhancing access to specialized expertise. Since its inception in 1986, when black-and-white pathology images were first transmitted via video [2], telepathology has rapidly advanced, benefiting from sophisticated imaging and telecommunication tools.

Telepathology systems can be categorized into static, dynamic, robotic, whole-slide imaging (WSI), and hybrid methods. In static telepathology, digital snapshots of specific areas are transmitted, though the quality of these snapshots is highly dependent on operator skill. Dynamic telepathology involves a live video feed from a microscope, enabling remote guidance in real time. Robotic microscopy enhances this process by allowing remote control over the microscope, eliminating the need for verbal instructions [3]. Whole-slide imaging (WSI), introduced in 2000, allows entire microscope slides to be digitized at high resolution [4]. This comprehensive approach supports analysis of a full range of pathology specimens, including H&E-stained paraffin sections, blood smears, and cytology slides [1]. WSI has become integral to telepathology, offering storage, archiving, and retrieval capabilities that traditional microscopy lacks. Hybrid systems combine aspects of robotic microscopy and WSI, and their use has expanded following the COVID-19 pandemic, which underscored the need for remote diagnostics and education.

Telepathology using WSI has potential to provide expertise in pathology subspecialties available in academic centers to laboratories around the United States and especially to other developing countries where there is a lack of subspecialists [5,6]. Numerous studies have demonstrated that WSI provides diagnostic accuracy comparable to traditional methods based on glass slides. Strong correlations have been documented in breast, gastrointestinal, pulmonary, and prostate biopsies, with WSI showing high concordance with glass-slide diagnoses [7]. One of the largest studies, by the University of Pittsburgh Medical Center and KingMed Diagnostics, reviewed 1,561 cases and established telepathology's reliability across specialties [8]. However, studies focusing exclusively on telepathology for hematopathology via WSI remain limited, particularly in resource-constrained settings.

**MATERIALS AND METHODS**
In May 2022, a telepathology initiative was established between the Pathology Department at Ho Chi Minh City Oncology Hospital (HCMCOH) in Vietnam and a hematopathologist (A.N.) at University of Texas Health-McGovern Medical School (UTHMMS) in Houston, Texas. The collaboration aimed to enhance hematopathology diagnostics through monthly virtual meetings where two to five challenging cases were reviewed for consultation.

All glass slides of telepathology cases were prepared and scanned at HCMCOH. WSIs of H&E and immunohistochemistry (IHC) stained slides, prepared from 3 µm sections, were scanned at 40x magnification using the VENTANA DP200 scanner (Roche, Basel, Switzerland) in 2022 (Figure 1) and the VS120 scanner (Olympus Corporation, Tokyo, Japan) from 2023 onwards (Figure 2). Prior to the telepathology conferences, these digital slides were uploaded to PathPresenter, a web-based platform facilitating high-resolution review of WSIs (Figure 3), together with other relevant data. Data included detailed clinical history, anatomical site information, and demographics, with all personal identifiers removed for confidentiality. The WSIs were posted online to be previewed by the hematopathologist at UTHMMS (typical WSIs in Figures 4 and 5). Virtual meeting was facilitated through Zoom Meeting (Zoom Video Communications, Inc., San Jose, California, U.S.), at 9PM (Houston time) corresponding to 9AM (Ho Chi Minh City time) due to different time zones (Figure 6). Discussion between Pathology Laboratory staff at HCMCOH and the hematopathologist at UTHMMS was carried out to get a consensus on the diagnosis for each case. If further data were needed, an action plan was suggested to obtain further test results for further discussion (in follow-up email or in follow-up meetings) for timely patient management. In select cases, additional IHC tests were recommended. When less common IHCs were unavailable locally, tissue blocks were sent to UTHMMS for additional staining (for a total of 5 IHC slides as of now in this project).

**RESULTS**
Over a 32-month period, 71 hematopathology cases were submitted for telepathology, covering patients with age ranging from 1 month to 83 years, a gender distribution of 36 males and 35 females. Various types of specimens were included in these telepathology sessions (Table 1). Specimens were primarily lymph nodes (43.2%) and bone marrow biopsies (31.8%). A definitive diagnosis was achieved in 97.2% of cases, with three cases remaining indeterminate due to either limited clinical information or insufficient tissue for further testing. Among cases reviewed, 74.6% were malignant, encompassing various hematologic disorders, including lymphomas, leukemias, and the remaining were non-malignant conditions (Table 2). The diagnoses for all cases are listed in Table 3.

Diagnostic efficiency improved with average turnaround times reduced by 30% compared to traditional on-site consultations with local pathologists using glass slides. Sensitivity and specificity of WSI-based diagnoses were recorded at 95% and 94%, respectively, underscoring the reliability of telepathology in accurately identifying malignant and non-malignant conditions.

A web site has been established in this telepathology project to retain information of the most recently discussed cases for further review by attendees after the teleconference.
https://hemepathreview.com/Nguyen/WSICaseList_Ver7.htm
From the main page of this website, a consultation case with a known diagnosis (after consensus in the virtual meeting) can be selected from the list (Figure 7). Upon going to the selected case, the following information will be displayed (Figure 8):
- Clinical history, including pertinent H&P findings, laboratory results and imaging findings
- PathPresenter links to WSIs which include H&E slides and IHC slides; some also include lab results and flow cytometry scattergrams
- Summary of histological findings that are found from the PathPresenter links above
- A diagnosis that has been obtained by consensus between HCMCOH staff and the hematopathologist at UTHMMS
- Suggestions as needed for further IHCs, laboratory tests, or more clinical data from oncology to clarify the diagnosis.

**DISCUSSION**
This study demonstrates the utility of telepathology, particularly through WSI, in enhancing hematopathology diagnostics in low-resource settings. The collaboration between HCMCOH and UTHMMS underscores telepathology's ability to bridge geographical and resource limitations in diagnostic expertise. Findings from this study are consistent with previous works demonstrating diagnostic concordance between WSI and traditional microscopy [7]. WSI provides an effective platform for real-time consultations, allowing remote subspecialized hematopathologists to interact with local pathologists for comprehensive review of challenging cases. This process also fosters ongoing education, facilitating knowledge transfer in regions where specialized hematopathology expertise is limited.

Despite promising results, limitations remain. Some regions may lack access to specific IHC stains, and in rare cases, physical transfer of tissue samples was necessary. Additionally, the high cost of telepathology equipment and infrastructure may restrict broader implementation in resource-limited settings [9]. Future advancements in telepathology must address these barriers to accessibility and sustainability.

Looking forward, the integration of artificial intelligence (AI) into telepathology workflows holds significant promise [10,11]. AI-supported image analysis could augment diagnostic accuracy, further enhancing telepathology's value in low-resource settings.

**ETHICS & COMPLIANCE**
All procedures were conducted following ethical guidelines, with institutional review board (IRB) approval obtained at both institutions. All case data were de-identified to maintain patient confidentiality.

Table 1. Specimen distribution across organ types

| Organ | | Specimen number | | | | |
|-------|--|------|------|------|-------|----------------|
| | | 2022 | 2023 | 2024 | Total | Percentage (%) |
| *Bone marrow* | | 4 | 13 | 11 | 28 | 31.8 |
| *Lymph node* | | 4 | 10 | 24 | 38 | 43.2 |
| *Spleen* | | | | 2 | 2 | 2.3 |
| *Skin* | | | 1 | 4 | 5 | 5.7 |
| *Bone* | *Mandible* | | 1 | | 1 | 1.1 |
| | *Maxilla* | | | 1 | 1 | 1.1 |
| | *Undefined* | | | 1 | 1 | 1.1 |
| *Head and neck* | *Nasopharynx* | | 1 | | 1 | 1.1 |
| | *Hard palate* | | 1 | | 1 | 1.1 |
| | *Nasal cavity* | | 1 | | 1 | 1.1 |
| | *Submandibular mass (salivary gland)* | | | 1 | 1 | 1.1 |
| *Gastrointestinal tract* | *Stomach* | | | 1 | 1 | 1.1 |
| | *Duodenum* | | | 1 | 1 | 1.1 |
| | *Colon-rectum* | | | 2 | 2 | 2.3 |
| *Breast* | | | | 1 | 1 | 1.1 |
| *Anterior chamber fluid* | | | | 1 | 1 | 1.1 |
| *Soft tissue* | *Chest wall* | | 1 | | 1 | 1.1 |
| | *Forearm* | | 1 | | 1 | 1.1 |
| *TOTAL* | | 8 | 30 | 50 | 88 | 100 |

Table 2. Distribution of malignant cases versus non-malignant cases

|  | **Case number** | **Percentage (%)** |
|---|---|---|
| *Benign* | 14 | 19.7 |
| *Borderline* | 2 | 2.8 |
| *Malignant* | 53 | 74.6 |
| *Undefined* | 2 | 2.8 |
| *TOTAL* | 71 | 100 |

Table 3. Diagnosis of cases in teleconferences

| **Diagnosis** | **Case number** |
|---|---|
| *Acute leukemia (ALL or AML)* | 1 |
| *Angioimmunoblastic T-cell lymphoma (AITL)* | 1 |
| *Anaplastic large cell lymphoma (ALCL)* | 4 |
| *Acute lymphoblastic leukemia (ALL)* | 2 |
| *Acute myeloid leukemia (AML)* | 1 |
| *Atypical plasma cell hyperplasia (cannot rule out solitary plasmacytoma or multiple myeloma)* | 1 |
| *B-cell lymphoma (TCR BCL or DLBCL)* | 1 |
| *CD5-positive diffuse large B-cell lymphoma (CD5-pos DLBCL)* | 1 |
| *Classic Hodgkin lymphoma (cHL)* | 4 |
| *Chronic lymphocytic leukemia/Small lymphocytic lymphoma (CLL/SLL)* | 2 |
| *Chronic myeloid leukemia (CML)* | 2 |
| *Dermatopathic lymphadenopathy* | 1 |
| *Disseminated juvenile xanthogranuloma* | 1 |
| *Follicular lymphoma (FL)* | 5 |
| *Follicular hyperplasia* | 5 |
| *ALK-positive histiocytosis* | 1 |
| *Hemophagocytic lymphohistiocytosis (HLH)* | 2 |
| *Hypocellular marrow, secondary to chemotherapy* | 2 |
| *Interdigitating dendritic cell sarcoma* | 2 |
| *Lymphoplasmacytic lymphoma* | 1 |
| *Multiple myeloma* | 2 |
| *Myeloid sarcoma* | 1 |
| *Myeloid sarcoma with AML* | 1 |
| *Marginal zone lymphoma (MZL)* | 1 |
| *Nodular lymphocyte predominant Hodgkin lymphoma (NLPHL)* | 1 |
| *Normal bone marrow* | 2 |
| *Normal bone marrow + Reactive polyclonal plasma cells (Rectum)* | 1 |
| *Osteomyelofibrosis (bone marrow) + Peripheral T-cell lymphoma, NOS (PTCL, NOS) (spleen)* | 1 |
| *Plasmablastic lymphoma (PBL)* | 2 |
| *Polymorphous adenocarcinoma* | 1 |
| *Primary cutaneous CD30-positive lymphoproliferative disorder* | 1 |
| *Peripheral T-cell lymphoma, NOS (PTCL, NOS)* | 4 |
| *Reactive lymphoid tissue* | 2 |
| *Refractory CD30-pos lymphoproliferative disorder* | 1 |
| *Splenic marginal zone lymphoma (SMZL)* | 1 |
| *T-cell lymphoma (CD8-positive)* | 2 |
| *T-lymphoblastic lymphoma* | 3 |
| *T-lymphoblastic lymphoma with T-ALL* | 1 |
| *Thalassemia* | 1 |
| *Undifferentiated acute leukemia* | 1 |
| *Undifferentiated carcinoma/sarcoma* | 1 |
| *TOTAL NUMBER OF CASES:* | 71 |

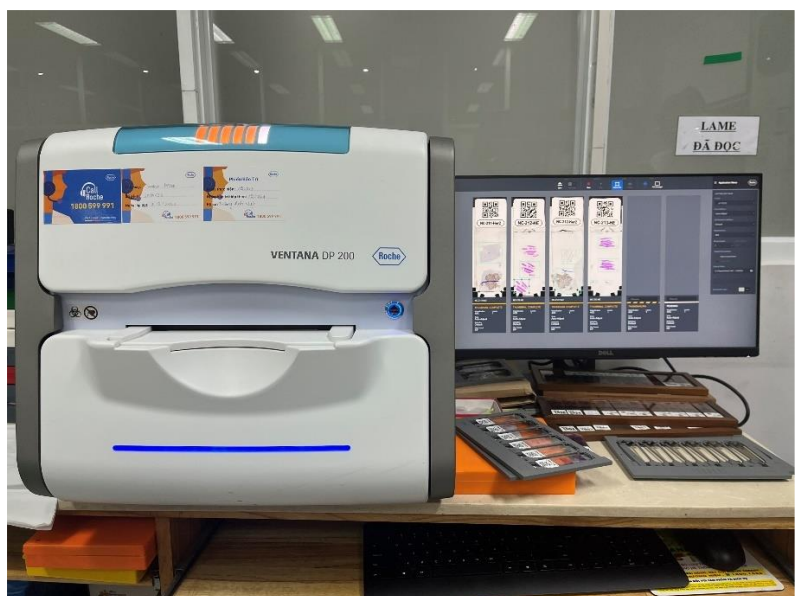

Figure 1. VENTANA DP 200 Slide Scanner at HCMC Oncology Hospital

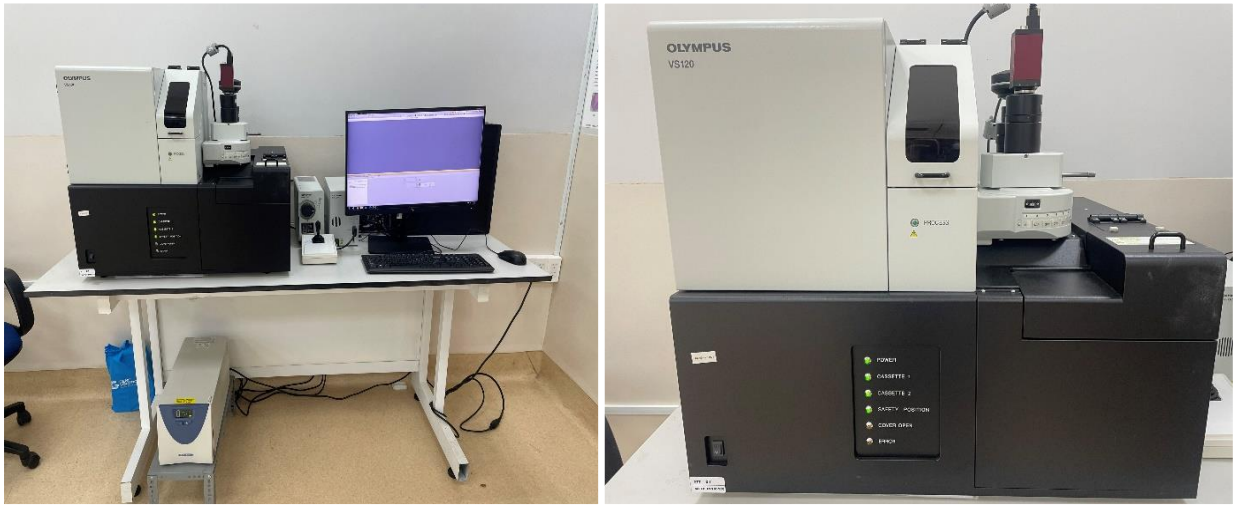

Figure 2. Olympus VS120 Virtual Slide Scanner at HCMC Oncology Hospital

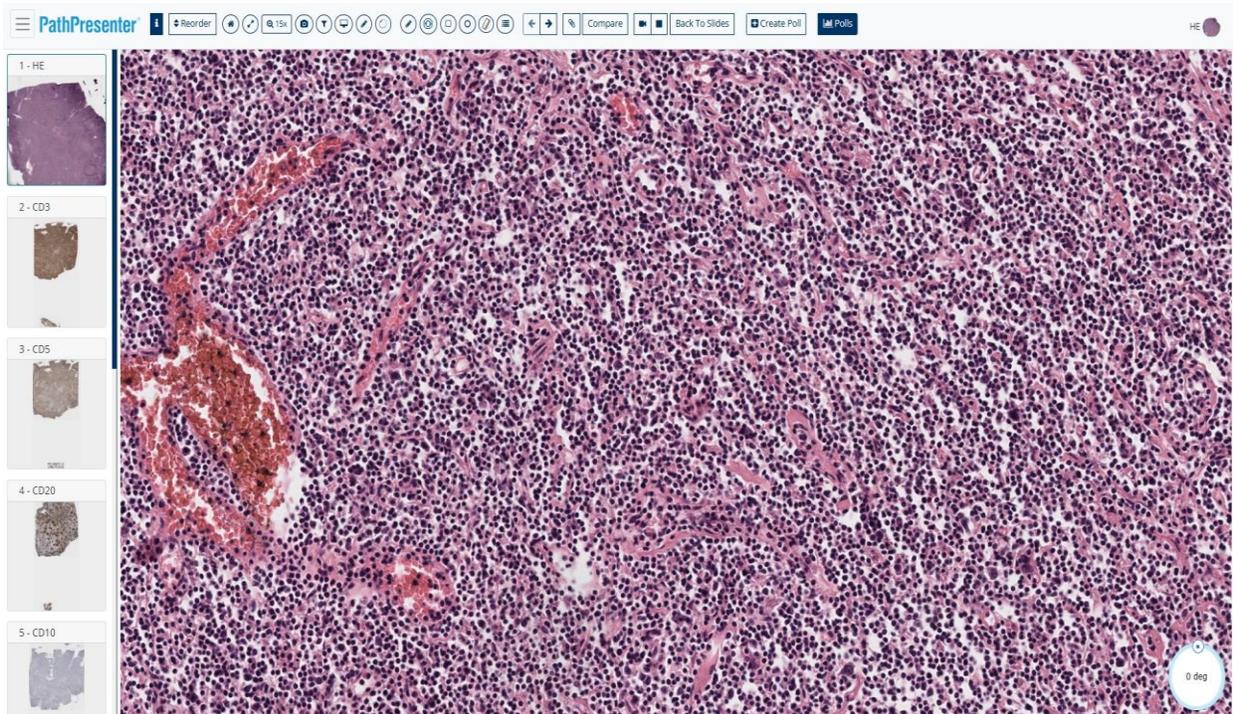

Figure 3. Screen view for full display of PathPresenter, a platform for sharing pathology images

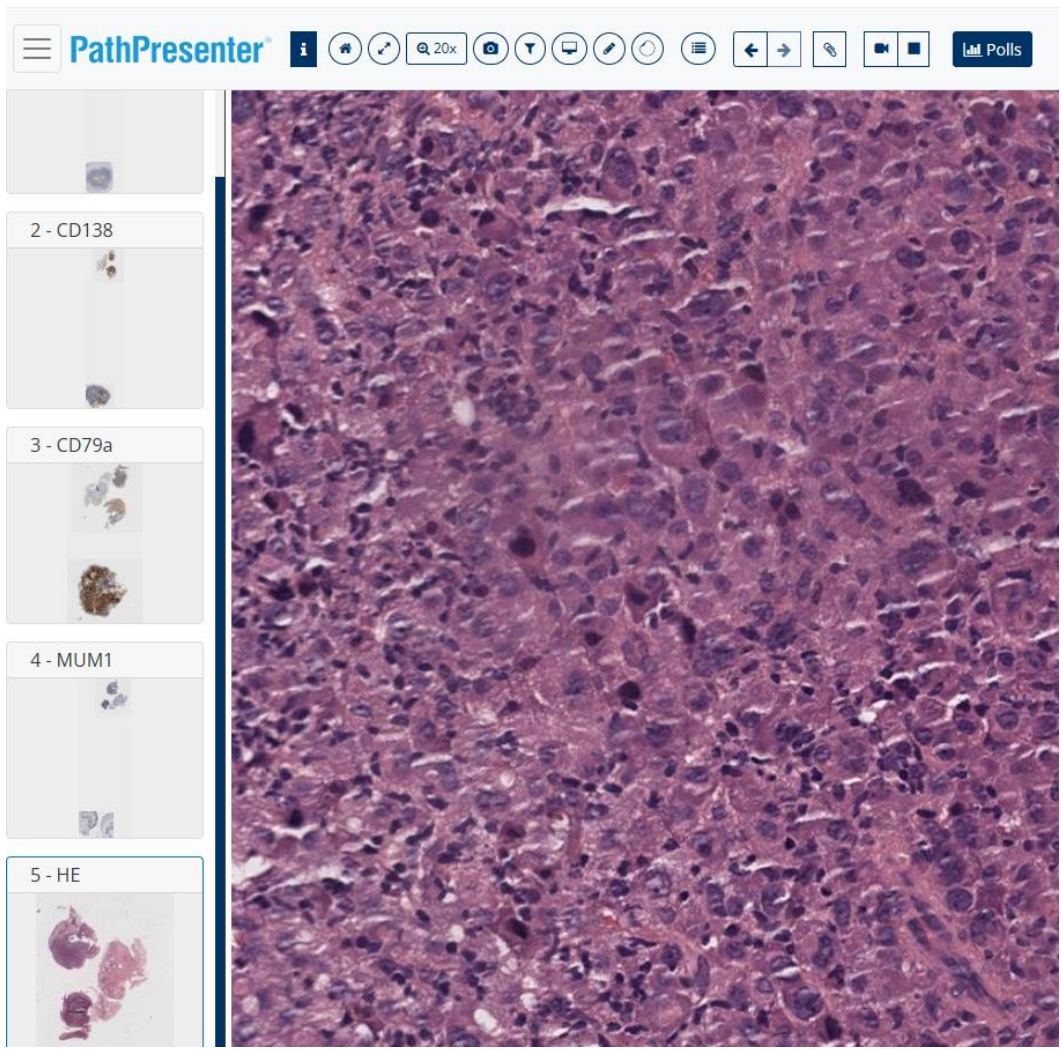

Figure 4. A representative WSI of H&E slide for a case of plasmablastic lymphoma

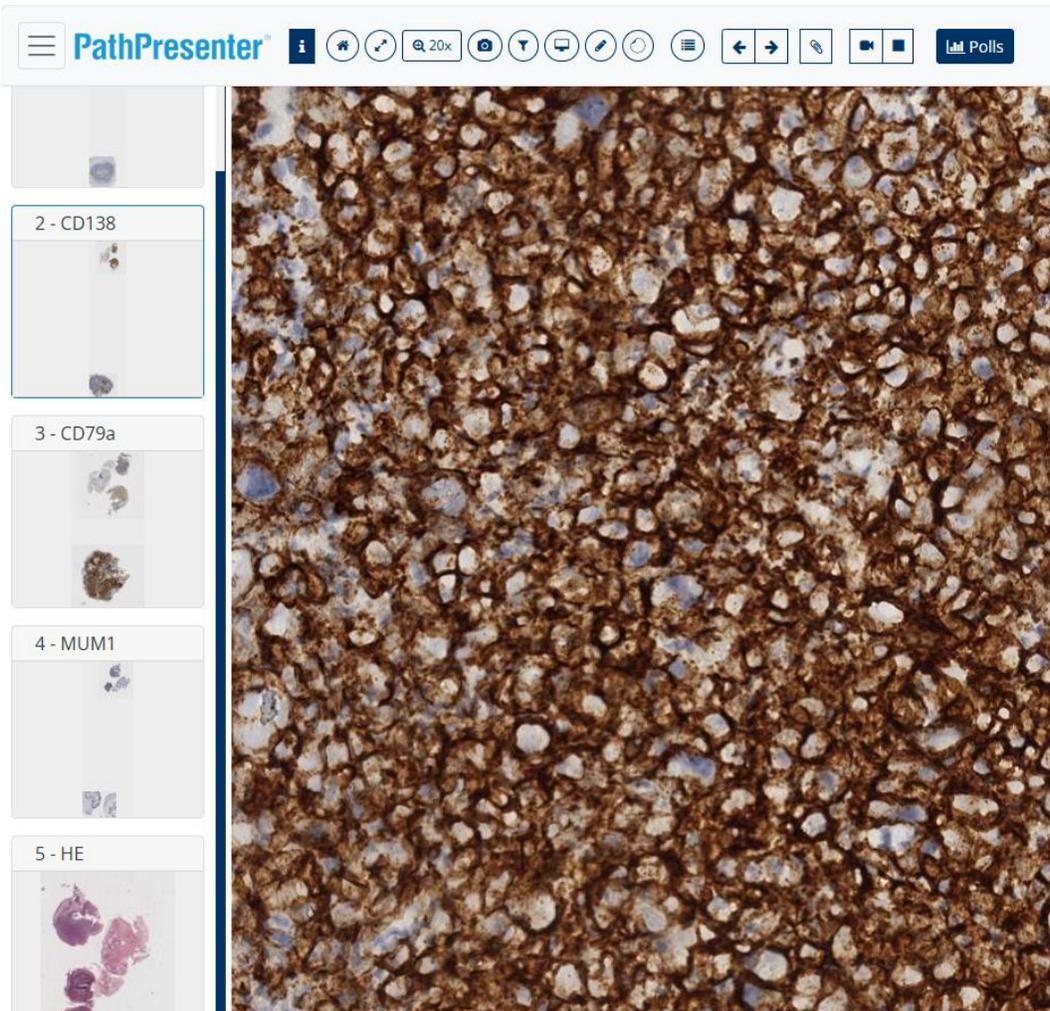

Figure 5. A representative WSI of IHC stain CD138 for a case of plasmablastic lymphoma

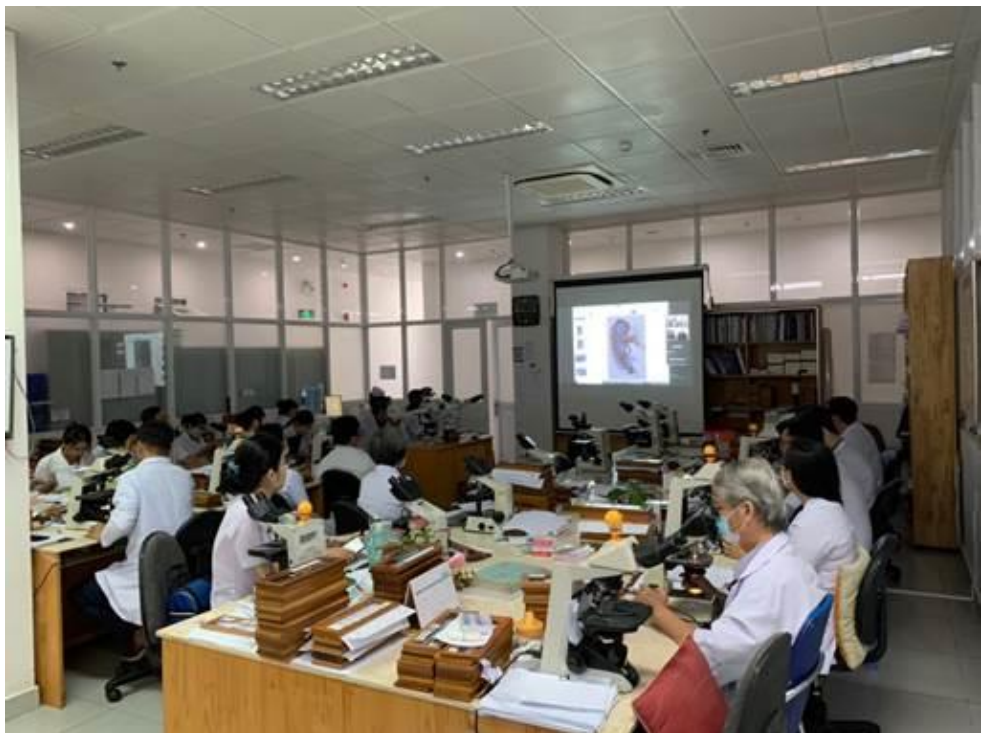

Figure 6. Pathology group at HCMC Oncology Hospital in the teleconference

**WHOLE-SLIDE-IMAGING (WSI) CASE LIST FOR TELEPATHOLOGY**
**Hematopathology cases at the Cancer Center, Ho Chi Minh City, Vietnam**
**Last Revision on: 11/15/2024**

**Uyen Ly, MD; Quang Nguyen, MD; Dang Nguyen, MD; Tu Thai, MD; Binh Le, MD; Duong Gion, MD; Alexander Banerjee, MD; Brenda Mai, MD; Amer Wahed, MD; Andy Nguyen, MD**

Case 19_6: 54 y/o male with likely undifferentiated acute leukemia
Case 20_1: 60 y/o male with primary cutaneous CD30-positive lymphoproliferative disorder
Case 20_2: 43 y/o female with anaplastic large cell lymphoma, ALK-negative
Case 20_3: 50 y/o male with CLL/SLL

Figure 7. Listing of WSI cases presented in recent teleconferences (partial list)
https://hemepathreview.com/Nguyen/WSICaseList_Ver7.htm

**CASE 3**:
50 y/o male, enlarged neck lymph node, CBC: WBC 211k, Lymphocyte 192K, Prolymphocyte 7%. BM Aspirate shows 40% lymphocyte.

Lymph node biopsy:
https://pathpresenter.net/public/presentation/display?token=0d9657cf

Findings:
Scattered HRS-like cells, background of small lymphocytes
HRS-like cells:
(+) CD30
(-) CD3, CD15, CD68
PAX5 out of focus
CD45: difficult to interpret (too crowded)
The HRS-like cells may be activated lymphocytes or true HRS cells
Suggestion: rescan PAX5 (to r/o cHL)

Small lymphocytes: include some CD3-pos cells
Suggestion: IHCs for: CD20, CD5, CD23 to r/o CLL/SLL

DX to consider:
-CLL/SLL (most likely)
-CLL/SLL together with cHL

BM Bx, Asp:
https://pathpresenter.net/public/presentation/display?token=9ab80383
Findings:
Large aggregates with small lymphocytes
(+)CD5, CD20, CD79a, PAX5, CD23, bcl2
(-) CD15, CD30, CD10, bcl6, TdT, MPO, CD117

Flow cytometry:
(+) CD5, CD19, CD20, CD79a, CD200, lambda
(-) CD10, CD34, CD38, CD3, CD7, CD4, CD8, kappa, CD2, CD56, TdT
Diagnosis: CLL/SLL
+++++

Figure 8. Summary of a representative teleconference case